# AI-driven Inverse Design of Complex Oxide Thin Films for Semiconductor Devices

Bonwook Gu[1], Trinh Ngoc Le[1], Wonjoong Kim[1], Zunair Masroor[1] and Han-Bo-Ram Lee[1*]

[1]Department of Materials Science and Engineering, Incheon National University, Incheon 22012, Korea

***Corresponding Authors**: hbrlee@inu.ac.kr

**Abstract**

Bridging generative foundation models with non-equilibrium thin-film synthesis remains a central challenge, limiting the practical impact of AI-driven materials discovery on semiconductor dielectrics. Here, we introduce IDEAL (Inverse Design for Experimental Atomic Layers), an inverse-design platform that links generative diffusion models, machine learning interatomic potentials, and graph neural network property predictors with atomic layer deposition (ALD). We demonstrate IDEAL using the Hf–Zr–O system as a stringent benchmark for semiconductor-relevant complex oxides. The platform statistically enumerates thermodynamically plausible structures and constructs a composition–structure–property map. Crucially, it identifies a narrow composition window where low-energy tetragonal and orthorhombic phases cluster, revealing trade-offs between band gap and dielectric response. Experimental validation using atomic layer modulation (ALM) corroborates these predictions, demonstrating predictive guidance under realistic, non-equilibrium thin-film growth. By experimentally closing the loop, IDEAL provides a transferable and generalizable route to the precision synthesis of next-generation semiconductor dielectrics.

# Introduction

Scaling semiconductor devices into complex three-dimensional architecture has significantly increased the demand for precise, conformal thin films. Atomic layer deposition (ALD) has become an essential technology to meet this demand, as it is one of the few methods capable of coating high-aspect-ratio structures with atomic-level precision.[1,2] However, extending ALD to multi-component systems creates a critical problem. Identifying the optimal elemental composition is extremely difficult. Since the ideal ratio is unknown beforehand, researchers are forced to rely on extensive trial-and-error to find suitable recipes. This trial-and-error paradigm fundamentally limits the pace at which new functional oxides can be integrated into advanced semiconductor devices. Furthermore, as systems expand to ternary compositions or beyond, the number of possible combinations increases drastically, making it practically impossible to experimentally explore the entire compositional space.

In contrast to these experimental limitations, recent breakthroughs in artificial intelligence (AI) have revolutionized material discovery. These technologies have advanced beyond simple property prediction to guide the actual synthesis of inorganic materials.[3,4,5] This evolution enables inverse-design strategies where functional targets dictate the material structure.[6,7] Foundation models like MatterGen have demonstrated this capability by successfully predicting stable bulk crystals.[8] Specifically, it has been utilized to propose new candidates for diverse applications, ranging from lithium-ion conductors to magnetic materials[5], by optimizing structures for low formation energy. These advances demonstrate strong potential for discovering thermodynamically plausible bulk materials, yet their relevance to non-equilibrium thin-film synthesis remains unclear. In parallel, machine learning models have increasingly been used to rapidly predict key materials properties from crystal structures, enabling high-throughput screening beyond what is feasible with first-principles calculations alone. These property predictors make it possible to map composition–structure–property relationships at scale and to prioritize candidates that balance stability with functional targets.[9, 10]

Despite these advances, a fundamental uncertainty remains regarding whether bulk-trained generative models can meaningfully guide thin-film synthesis. Foundation models like MatterGen are primarily designed to predict stable bulk crystals.[8] Most prior MatterGen-based studies follow a screening-driven paradigm, in which a limited number of candidate materials are generated, the most promising ones are selected based on predicted properties, and their feasibility is subsequently validated through experiments. In contrast, semiconductor devices rely almost exclusively on thin films, where phase formation is governed not only by bulk thermodynamics but also by surface and interface energies under highly non-equilibrium processing conditions.[11] Moreover, practical semiconductor integration severely constrains the accessible material space, making it far more challenging to identify optimal compositions and crystal structures within a predefined elemental system rather than discovering entirely new compounds. Consequently, it remains unverified whether bulk-optimized generative predictions can be translated to the complex, non-equilibrium environment of thin-film growth.

To address this gap, we move beyond conventional candidate selection strategies and adopt a fully statistical framework. Specifically, we use MatterGen to enumerate all plausible crystal structures within a known elemental combination and subsequently evaluate their composition-dependent physical properties using a graph neural network (GNN)-based model. This comprehensive approach enables systematic exploration of the composition-dependent property landscape rather than isolated material selection. Finally, we experimentally validate whether the predicted compositions and phases can be realized through ALD, thereby establishing a direct experimental link between a generative diffusion model and ALD, the industrial standard for precision thin-film manufacturing.

In this work, we introduce IDEAL (Inverse Design for Experimental Atomic Layers), an inverse-design platform that operationalizes this approach as shown in Figure 1. We select the widely studied Hf-Zr-O family, specifically $Hf_{1-x}Zr_xO_2$, as a robust test bed to demonstrate the platform's efficacy. IDEAL combines generative structure sampling, rapid relaxation and stability analysis, and property prediction to construct a composition–structure–property map for experiment. We then validate the predicted composition and phase window through atomic layer modulation (ALM) based on ALD as shown in Figure S1, testing whether bulk-trained generative outputs remain predictive under non-equilibrium thin-film growth. The resulting thin films corroborate the model's predictions, confirming that IDEAL can efficiently guide the synthesis of complex oxides with minimized trial-and-error. Because the workflow is modular, the pretrained models can be readily swapped as improved generative models, interatomic potentials, or property predictors emerge, enabling continuous upgrades in accuracy without changing the overall design logic.

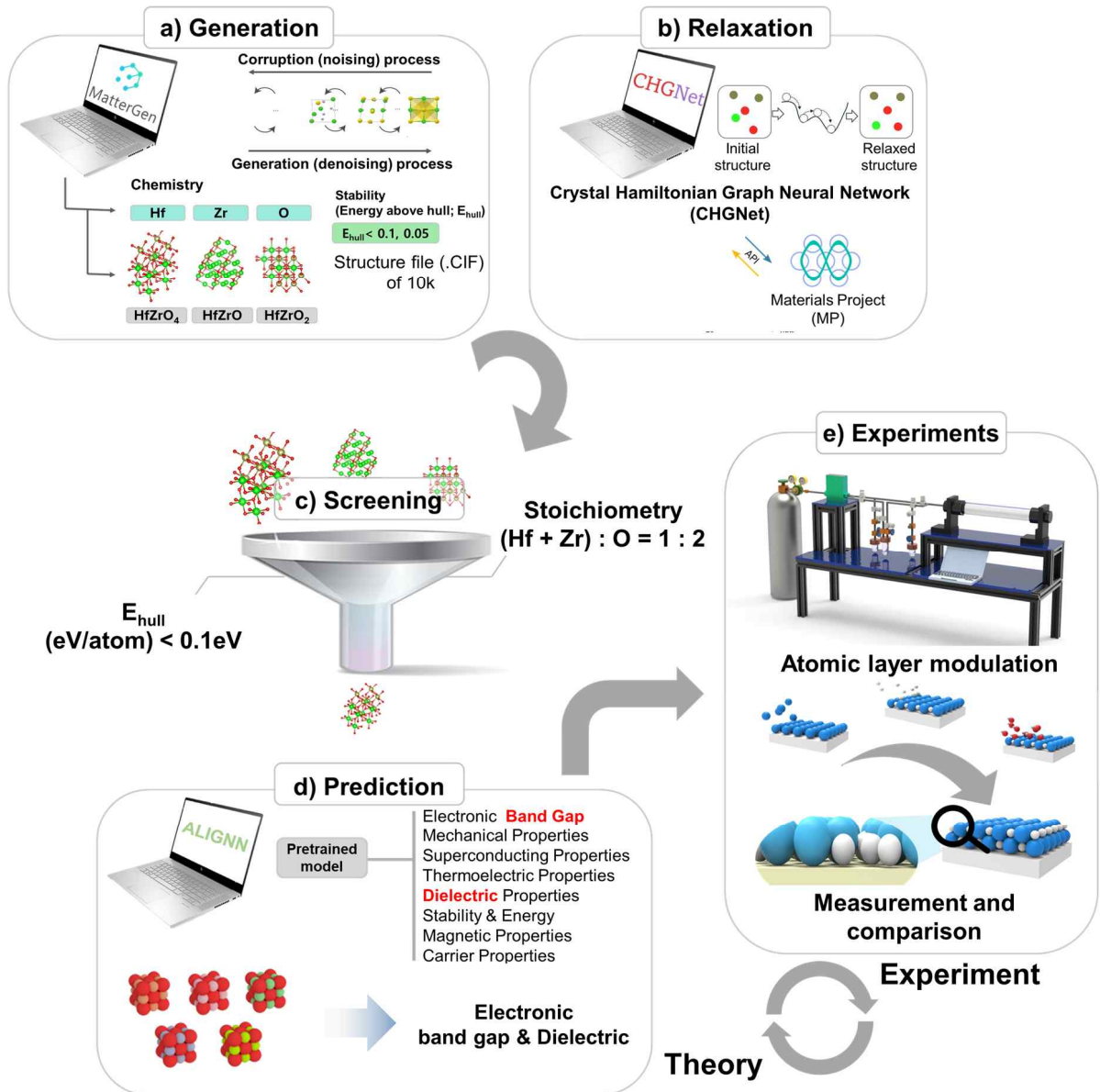

Figure 1. Overall workflow of the IDEAL platform, which connects a) MatterGen based structure generation, b) CHGNet based relaxation and stability analysis, c) screening, d) ALIGNN based property prediction, and e) ALM experiments in the Hf–Zr–O system.

## Results and Discussion

**Validation of CHGNet and ALIGNN models**

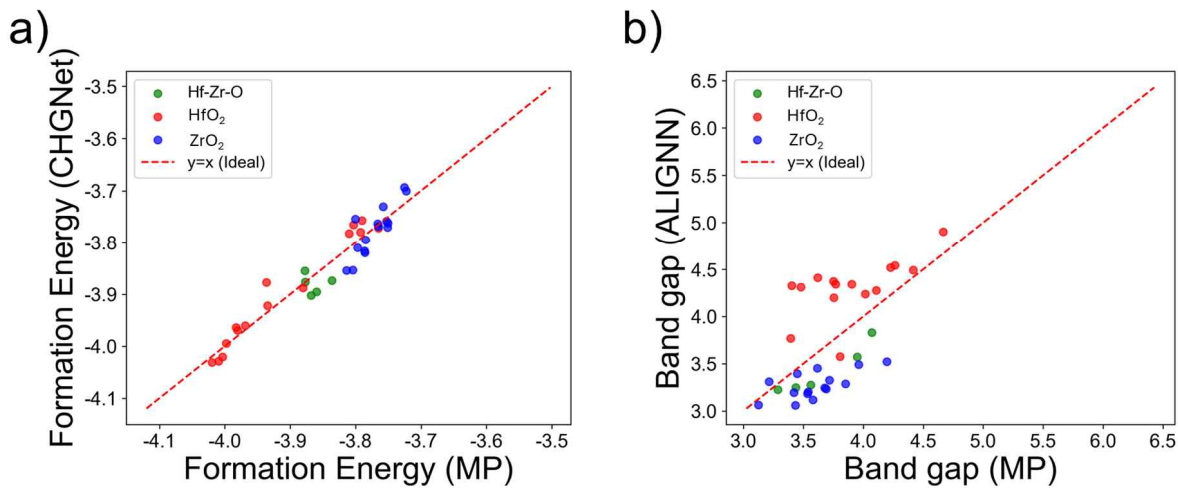

Figure 2. Benchmark of CHGNet and ALIGNN against the Materials Project (MP) database for selected $HfO_2$, $ZrO_2$, and $Hf_{0.5}Zr_{0.5}O_2$ structures, demonstrating a) formation energies from CHGNet and b) band gap from ALIGNN.

Before applying the generative platform to the large set of MatterGen structures, we rigorously assessed the predictive fidelity of CHGNet (Crystal Hamiltonian Graph Neural Network) and ALIGNN (Atomistic Line Graph Neural Network) against standard DFT benchmarks. We assembled a reference set of 32 structures from the Materials Project spanning the Hf–O, Zr–O, and Hf–Zr–O chemical systems as shown in Table S1. These structures correspond to the stable and metastable reference phases used for the convex-hull construction in our self-consistent screening (Section 2.4). These structures include the relevant oxide polymorphs and chemical environments needed to benchmark energies and electronic properties for the $Hf_{1-x}Zr_xO_2$ family. For benchmarking, we compared the Materials Project-reported values against the predictions made by CHGNet and ALIGNN using the same crystal structures as input.

Figure 2a shows that CHGNet formation energies exhibit exceptional agreement with DFT-computed values, with a coefficient of determination ($R^2$) of 0.915 and a Spearman rank correlation of 0.919, indicating that the model largely preserves the ranking across structures even when absolute values are biased. Furthermore, the mean absolute error (MAE) is remarkably low at 0.022 eV atom⁻¹, which is comparable to the intrinsic error margin of DFT methods themselves.[12] Although we employed a strictly self-consistent CHGNet-based $E_{hull}$ for the actual screening process (as detailed in Section 2.4) to eliminate systematic errors, this independent validation confirms that the underlying model physics is robust.[13]

For electronic properties, the ALIGNN band gap predictions were evaluated against the Materials Project benchmarks. The model exhibits a moderate correlation with DFT data ($R^2$ = 0.502) and a Spearman rank correlation of 0.636, with a MAE of 0.365 eV, as shown in Figure 2b. While this level of accuracy does not replace high-precision calculations, ALIGNN successfully captures the systematic trends across different compositions and structures. In the context of high-throughput screening, where the primary goal is to rank candidates rather than predict the exact band gap, this performance is sufficient.[14,15] We therefore utilize ALIGNN as a statistical screening tool, relying on aggregated trends by hundreds of candidates to guide material selection while averaging out individual stochastic errors.

Validating the dielectric response presents a unique challenge. While the total dielectric constant is the primary figure of merit for device applications, the prediction of the ionic contribution to the dielectric constant via machine learning remains a significant bottleneck.[16] This is because the ionic contribution involves complex lattice vibrational modes, which require computationally expensive density functional perturbation theory (DFPT) calculations.[17][18] This high computational cost leads to a severe scarcity of high-fidelity ground-truth data. For instance, statistical data from the JARVIS-DFT database reveals that while frequency-dependent dielectric tensors (electronic contribution) are available for over 34,000 structures, full DFPT-calculated dielectric tensors (including ionic contribution) cover only about 5,000 entries[19]. Consequently, machine learning models for the total dielectric response face higher uncertainty due to limited training data. In contrast, the electronic dielectric constant depends primarily on the frozen-ion electronic band structure.[20] Consequently, it benefits from larger training datasets and is predicted to have superior relative accuracy compared to the total dielectric constant by ALIGNN. Crucially, we justify the use of the electronic dielectric constant as a screening proxy. Although the ionic contribution constitutes a significant portion of the total permittivity in high-k oxides, previous high-throughput studies demonstrate a strong correlation between the electronic and total dielectric constants across diverse crystal systems.[7, 21] Specifically, in the Hf–Zr–O system, first-principles calculations confirm that the transition from the monoclinic phase to higher-symmetry tetragonal or cubic phases is accompanied by a simultaneous increase in both electronic and ionic polarizabilities.[22, 23] This occurs because the phase transformation involves densification and symmetry elevation, which enhances both the electronic susceptibility and the contribution from soft phonon modes.[7, 23, 24, 31, 32] Therefore, even without explicit phonon calculations, the electronic dielectric constant serves as a robust physical descriptor for identifying composition trends that favor high-k phases.

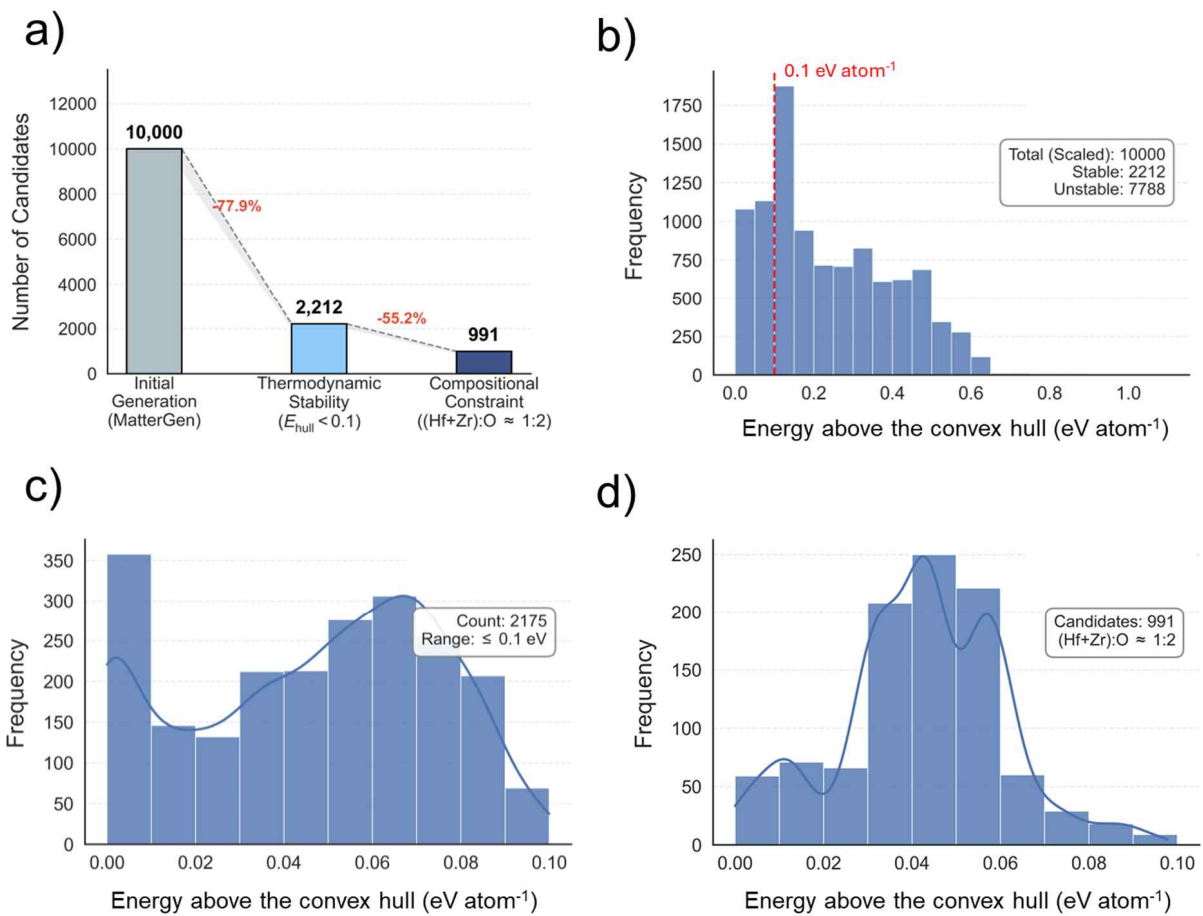

Figure 3. (a) Overview of the filtering workflow. Distributions of energy above the convex hull ($E_{hull}$) for (b) all MatterGen-generated Hf, Zr, and O structures, (c) the subset after thermodynamic stability filtering ($E_{hull} < 0.1$ eV atom$^{-1}$), and (d) the final subset after stoichiometry filtering.

**Thermodynamic filtering of generative candidates**

We first examine how the generative sampling and screening stages of the platform narrow down the search space in the Hf–Zr–O system. MatterGen was used to sample 10,000 hypothetical Hf–Zr–O crystal structures under the stability prior described in Section 2.2. These samples span a broad range of Hf-to-Zr ratios and include a variety of coordination environments and lattice symmetries, without any explicit constraint on the cation-to-oxygen ratio at the generation stage as shown in Figure 3a.

After full cell relaxation with CHGNet, the formation energy of each generated structure was evaluated, and its $E_{hull}$ was computed relative to the self-consistent Hf–Zr–O convex hull, as described in Section 2.4. The resulting distribution of $E_{hull}$ for all 10,000 candidates is shown in Figure 3b. Most structures lie between zero and about 0.6 eV atom$^{-1}$ above the hull, with a long tail extending to higher energy. The vertical dashed line in Figure 3b marks the threshold of 0.1 eV atom$^{-1}$ that we adopt as a practical criterion for thermodynamic plausibility. Applying this filter reduces the set from 10,000 down to 2,212 candidates, while 7,788 structures are discarded as energetically unfavorable, as shown in Figure 3c and Figure S2. Thus, only about 22% of the generative samples possess an $E_{hull}$ within 0.1 eV atom$^{-1}$ and are retained for further analysis. This discrepancy between the generation objective and the validated stability is expected for two reasons. First, the diffusion model provides a probabilistic prior rather than enforcing a hard constraint, so generated structures are not guaranteed to remain within the target $E_{hull}$ after full relaxation.[8] Second, our screening evaluates stability using a self-consistent CHGNet framework, in which both the generated candidates and the reference phases used to construct $E_{hull}$ are relaxed and evaluated on the same CHGNet potential-energy surface. Because this CHGNet-based $E_{hull}$ is not identical to the DFT-based stability information implicit in the training data and conditioning signal, candidates sampled near the target can shift above or below the threshold after CHGNet relaxation and $E_{hull}$ reconstruction.[10]

Although these 2,212 structures are thermodynamically closer to the convex hull, their Hf, Zr, and O ratios still include compositions unlikely to correspond to realistic oxide films, such as strongly oxygen-deficient or over-oxidized stoichiometries. To focus on compositions corresponding to electrically stable and charge-neutral $Hf_{1-x}Zr_xO_2$ phases, we imposed an additional stoichiometric window. Structures whose cation-to-oxygen ratio is approximately 1:2 were selected. This compositional filter minimizes the presence of metallic suboxides or defect-prone phases, reducing the set from 2,212 to 991 structures, as shown in Figure 3d and Figure S2. These remaining candidates are both thermodynamically stable and chemically consistent with the $Hf_{1-x}Zr_xO_2$ stoichiometry, corresponding to approximately 10 % of the original MatterGen samples. The distribution of $E_{hull}$ for these 991 $Hf_{1-x}Zr_xO_2$-like candidates is shown in Figure 3d. The resulting ensemble provides a manageable yet diverse set of plausible candidates to analyze structure-property relationships with ALIGNN.

**Consistency of ALIGNN predictions with reference data**

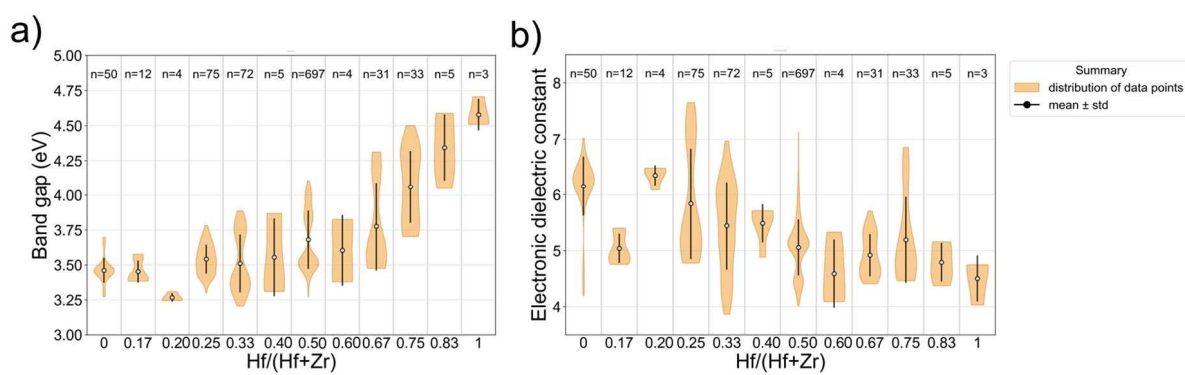

Figure 4. ALIGNN predicted a) band gap and b) electronic dielectric constant for 991 relaxed $Hf_{1-x}Zr_xO_2$ like structures as a function of Hf/(Hf+Zr).

Before using the ALIGNN predictions to guide process design, we first examined whether the predicted composition trends are consistent with available theory and experiments. Figure 4 shows the ALIGNN predicted band gap and electronic dielectric constant for 991 relaxed $Hf_{1-x}Zr_xO_2$-like structures that satisfy an $E_{hull}$ smaller than 0.1 eV atom$^{-1}$. Each point corresponds to one relaxed structure and is plotted as a function of Hf/(Hf+Zr). As Hf/(Hf+Zr) increases, the predicted band gap increases in Figure 4a, whereas the electronic dielectric constant decreases in Figure 4b. Zr-rich compositions, therefore, combine a relatively large electronic dielectric constant with a smaller band gap, and Hf-rich compositions show the opposite behavior.

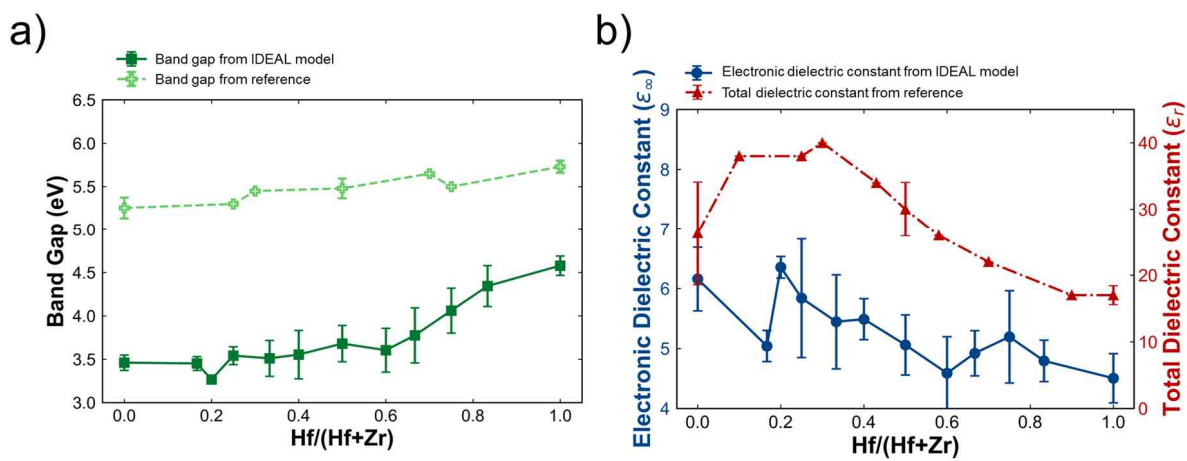

Figure 5. Comparison of a) ALIGNN band gap and b) electronic dielectric constant trends with experimental reference data for $Hf_{1-x}Zr_xO_2$.

To evaluate how realistic these composition trends are, Figure 5 compares the ALIGNN-predicted band gaps with reference band gaps and static dielectric constants. Detailed reference values are provided in Table S2. Since $Hf_{1-x}Zr_xO_2$ thin films often contain multiple polymorphs that vary with annealing, stress, and interfaces, we avoid selecting a specific crystal system and assess composition-dependent trends using all predicted structures. Accordingly, we focus on composition-dependent averages and spreads rather than assigning a single phase-specific value at each composition. In Figure 5a, the band gap is plotted as a function of Hf/(Hf+Zr) for two datasets: composition-dependent band gaps from the ALIGNN prediction (dark green symbols with error bars), and experimental optical band gaps for $Hf_{1-x}Zr_xO_2$ films (lime green diamonds). All curves show an increase in the band gap with increasing Hf content, and the slope of the ALIGNN curve is similar to that of the experimental reference curves. The absolute band gaps from ALIGNN are smaller than the experimental values, mainly because the pre-trained ALIGNN model was trained on DFT data. For many wide-band-gap oxides, the band gaps obtained from DFT are smaller than experimental values by roughly 1–2 eV, so DFT-based training data systematically underestimate the true band gaps [25].

Figure 5b compares the electronic dielectric constant predicted by ALIGNN with reference values. The dark blue dashed line represents the reference electronic dielectric constant, while the red dashed line corresponds to the reference static dielectric constant. Although the predicted electronic dielectric constant is smaller in magnitude than the total dielectric constant due to the exclusion of ionic contributions, its compositional dependence closely follows the reference trends. It is important to note that while the magnitude differs due to the lack of ionic contribution, the ALIGNN model accurately captures the relative ordering of phases and the composition where the dielectric response is maximized. This agreement confirms that the electronic dielectric constant effectively captures the phase-dependent polarizability changes in the Hf–Zr–O system, serving as a reliable indicator for the total dielectric response. In all three datasets, the dielectric constant decreases almost linearly as the Hf ratio increases, indicating that Zr-rich compositions exhibit the highest dielectric constants.

Taken together, Figures 4 and 5 show that the ALIGNN of IDEAL platform predictions provide a physically meaningful description of the composition trends in $Hf_{1-x}Zr_xO_2$. The model underestimates the absolute band gap and dielectric constant relative to experiment because it inherits the bias of the DFT training data.[26,27] Even so, it reproduces the monotonic change with Hf/(Hf+Zr) and the balance between a larger band gap at Hf-rich compositions and a larger dielectric constant at Zr-rich compositions. This level of agreement indicates that the ALIGNN dataset is sufficiently reliable to use as a screening tool in the following section, where it is combined with thermodynamic stability information to define a practical composition and phase window for process design.

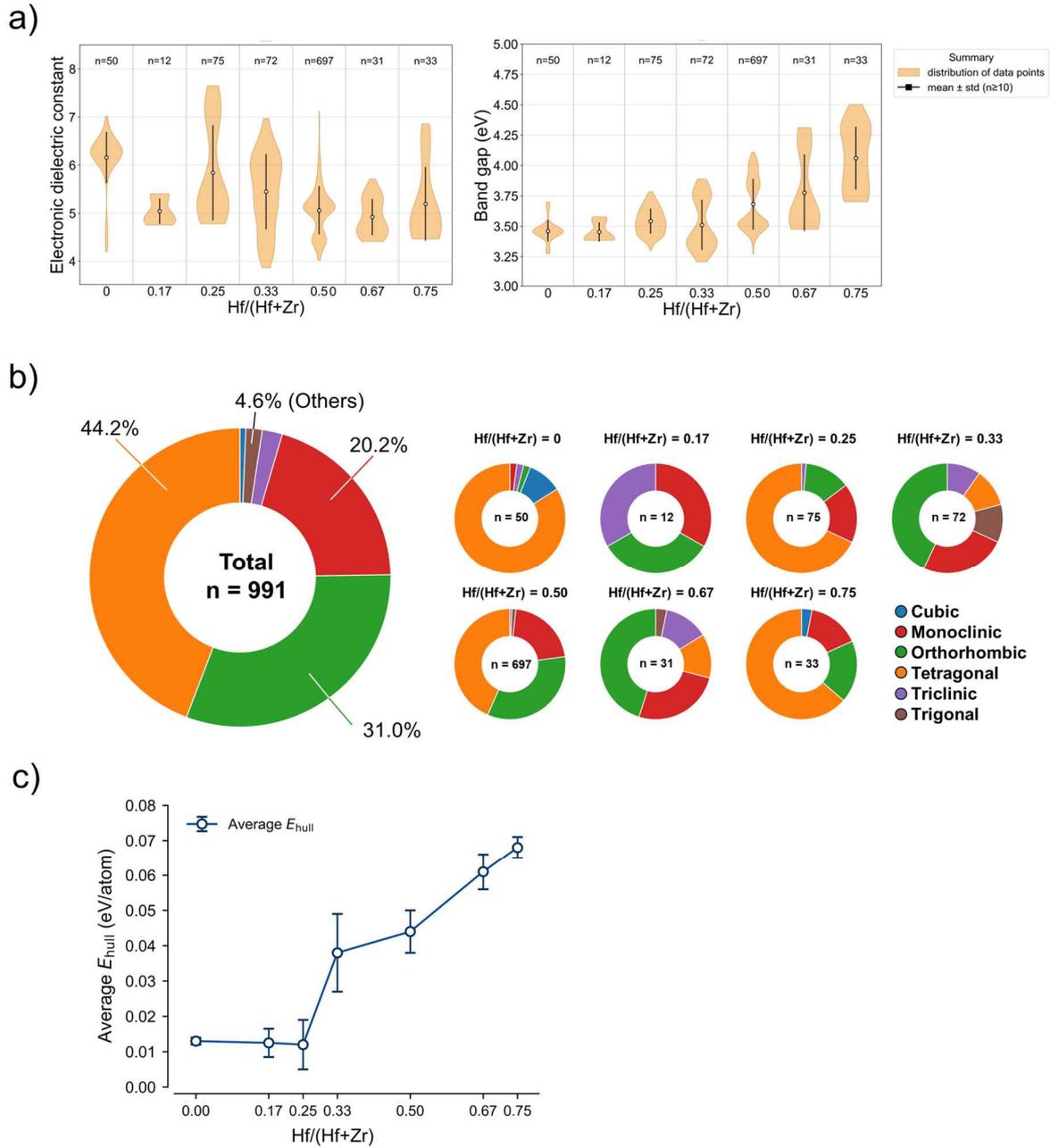

Figure 6. (a) Violin plots of the electronic dielectric constant and band gap versus Hf/(Hf+Zr), shown only for composition bins with n > 10; black markers indicate the mean and standard deviation. (b) Overall crystal system fractions of the 991 candidates and composition resolved crystal system fractions. (c) Average energy above the convex hull ($E_{hull}$) as a function of composition.

## Model-guided selection of composition and phase window

Having confirmed the reliability of the ALIGNN predictions in Section 3.2, we now define a practical composition window for device applications by analyzing the interplay between functional properties, crystal symmetries, and thermodynamic stability. To ensure statistical reliability, we restricted the analysis to composition bins containing more than 10 relaxed structures (n > 10), corresponding to Hf/(Hf+Zr) = 0, 0.17, 0.25, 0.33, 0.50, 0.67, and 0.75 (n = 50, 12, 75, 72, 697, 31, and 33, respectively) as shown in Figure 6a. The strong peak at Hf/(Hf+Zr) = 0.50 means that trends near the equimolar composition are supported by the largest statistical weight, whereas bins with smaller n should be interpreted with more caution. We attribute this strong accumulation of candidates near the equimolar composition to the inherent bias in the training data. As reflected in the Materials Project-derived reference set (Table S1), known ternary structures in the Hf–Zr–O system are predominantly concentrated at the stoichiometric 1:1 ratio ($Hf_{0.5}Zr_{0.5}O_2$). Consequently, the generative model, which learns to approximate the distribution of stable materials, reflects this data availability by preferentially generating structures near this specific composition.[28] However, this context underscores the robustness of our inverse design platform. By successfully validating the property trends across the full compositional range, including the data-sparse Hf-rich and Zr-rich regions against experimental results, we demonstrate that the model captures the intrinsic material physics rather than merely reproducing the training data distribution. Crucially, the violin plots in Figure 6a reveal a significant variation in properties at each composition. From a device perspective, this spread is important because it implies that composition alone does not uniquely determine the electronic response, and that phase selection and processing will control which portion of the distribution is realized in practice. Consistent with the overall trend, Zr-rich compositions (for example, Hf/(Hf+Zr) < 0.5) favor larger electronic dielectric constants but smaller band gaps, while Hf-rich compositions (for example, Hf/(Hf+Zr) > 0.5) move toward larger band gaps with reduced electronic dielectric response.

Figure 6b breaks down the generated structures by crystal system. Overall, the dataset is dominated by tetragonal (44.2%) and orthorhombic (31.0%) phases, followed by monoclinic (20.2%), with the remaining symmetries contributing 4.6% in total. Because the dataset is heavily concentrated at Hf/(Hf+Zr) = 0.50 (n = 697), these overall fractions are strongly influenced by the intermediate composition regime. The composition resolved distributions show that the equimolar composition is primarily composed of tetragonal and orthorhombic structures, which are the technologically relevant phases for high-k and ferroelectric applications [29] [30] [31]. This qualitative tendency is consistent with widely reported phase behavior in $Hf_{0.5}Zr_{0.5}O_2$ thin films, where the 1:1 composition most frequently yields tetragonal and orthorhombic signatures under processing conditions[31]. A practical implication is that the intermediate composition region is not only attractive for its property balance in Figure 6a, but also because it naturally populates the two phases that dominate device discussions.

Thermodynamic stability is summarized in Figure 6c, which plots the average $E_{hull}$ for each composition. The average $E_{hull}$ remains low for Zr-rich compositions up to Hf/(Hf+Zr) = 0.25, then increases sharply at Hf/(Hf+Zr) = 0.33 and continues to rise toward the Hf-rich end. This trend indicates that, within the relaxed structure set produced by the pipeline, compositions with higher Hf content are progressively less likely to yield low-energy polymorphs, and that maintaining thermodynamic favorability becomes more challenging as the composition moves into the Hf-rich regime. Importantly, this stability trend provides an independent lens that helps interpret the property distributions in Figure 6a. Although Hf-rich compositions tend to increase the band gap, the candidate structures in this region have higher average $E_{hull}$, indicating reduced thermodynamic favorability[32].

Combining these insights yields a clear design guideline without relying on a single hard threshold. Hf-rich compositions shift the electronic properties in a favorable direction for leakage control by widening the band gap, but they also move into a region of reduced thermodynamic favorability and a smaller pool of low-energy candidates. Zr-rich compositions are comparatively stable and provide larger electronic dielectric constants, but they are limited by narrower band gaps. Therefore, the intermediate composition range, roughly Hf/(Hf+Zr) between about 0.30 and 0.50, emerges as the most practical window because it simultaneously (i) preserves the band gap versus dielectric constant trade-off in a balanced regime, (ii) concentrates the phase statistics into tetragonal and orthorhombic structures, and (iii) avoids the strong increase in average $E_{hull}$ observed on the Hf rich side.

$E_{hull}$ is a bulk equilibrium metric and does not capture thin-film stabilization mechanisms such as strain, surface and interface energies, grain-size effects, and kinetic trapping. Nevertheless, it serves as a practical first-pass constraint that removes highly implausible candidates and focuses the search on metastable structures that are energetically accessible under realistic processing.

Finally, we emphasize why incorporating a thermodynamic filter is meaningful for thin film design and why it is useful in this work. In real ALD-derived oxide films, the observed phases reflect a competition between thermodynamic driving forces and kinetic or processing stabilization, including annealing temperature, stress, interfaces, and grain size[32, 31]. Consequently, thin films can realize metastable polymorphs that would be unlikely as bulk equilibrium phases, often facilitated by interface confinement or surface effects. However, candidates with excessively high formation energy are impossible to observe because their intrinsic instability imposes prohibitive barriers to nucleation and retention during thermal processing.[33] In this context, our thermodynamic filtering based on $E_{hull}$ effectively identifies the energetic window where these metastable phases are physically accessible, distinguishing them from highly unstable structures that cannot be realized even with the aid of thin-film effects. To our knowledge, there have been no prior systematic comparisons between thermodynamic filtering of generated polymorph sets and experimentally reported phase trends in inorganic thin films using generative models. In this study, when we account for thermodynamic favorability through $E_{hull}$-based filtering and interpret the results jointly with the phase statistics, the resulting composition window and the prominence of tetragonal and orthorhombic phases near Hf/(Hf+Zr) = 0.5 are consistent with the experimentally emphasized regime in $Hf_{0.5}Zr_{0.5}O_2$ thin films. This agreement supports the use of thermodynamic considerations as a practical constraint when translating generative predictions into experimentally actionable ALD design targets.

**Experimental implementation and validation of the model-guided design window**

Section 3.2 showed that the ALIGNN- and CHGNet-based screening narrows the $Hf_{1-x}Zr_xO_2$ design space to an intermediate window with Hf/(Hf+Zr) between about 0.30 and 0.50. In this range, ALIGNN predicts a clear trade-off: Hf-rich compositions have a larger band gap and smaller dielectric constant, whereas Zr-rich compositions have a smaller band gap and larger dielectric constant. The thermodynamic analysis in Figure 6c further indicates that low-energy tetragonal and orthorhombic structures cluster near Hf/(Hf+Zr) ≈ 0.50 and are rare at more Hf-rich or Zr-rich compositions.

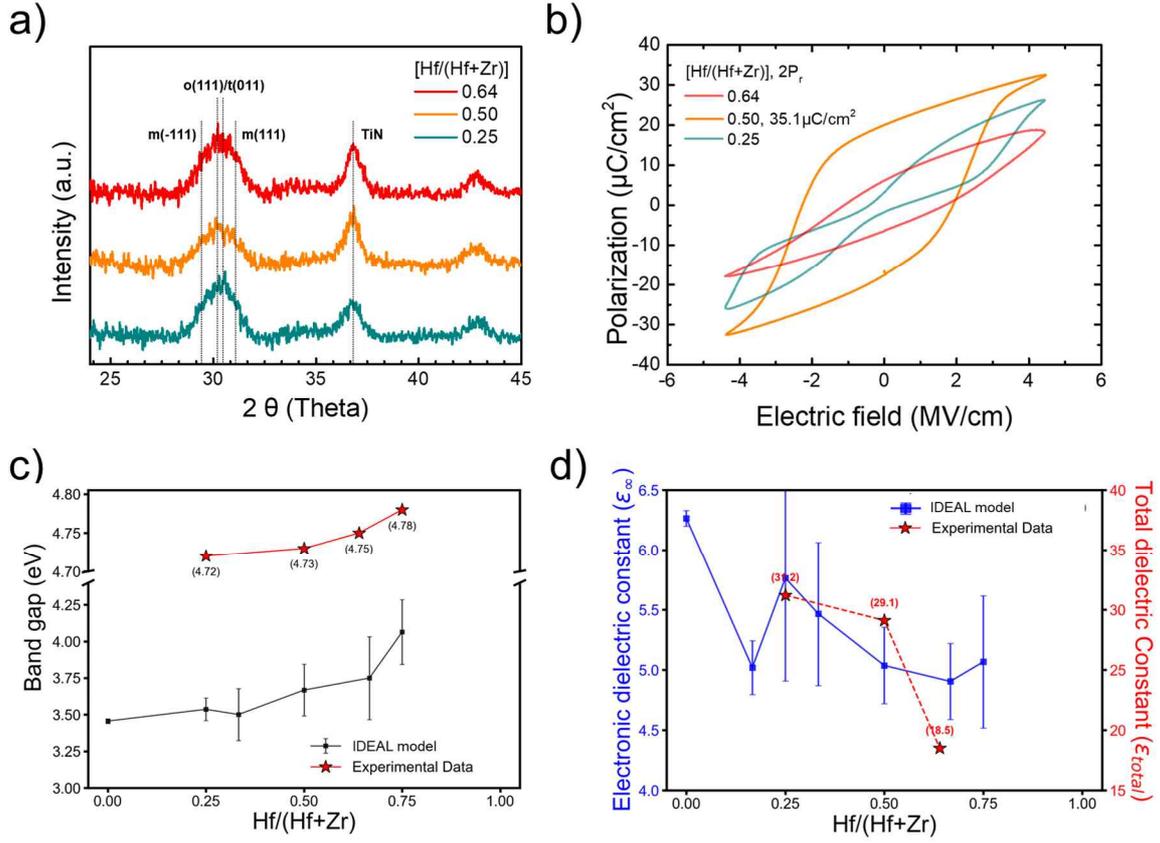

Figure 7. Experimentally measured a) Grazing-incidence X-ray diffraction, b) polarization–electric field curves reproduced from our previous ALM study[34], c) optical band gap, and d) total dielectric constant of ALM grown $Hf_{1-x}Zr_xO_2$ films guided by the IDEAL model.

To test this IDEAL window, we deposited $Hf_{1-x}Zr_xO_2$ films using the ALM process described in Section 2.7. The films were grown by repeating a [Zr–Hf] supercycle consisting of TEMAZ, purge, TDMAH, purge, and $O_3$ under identical growth conditions. The thickness was adjusted to about 4.5 nm. We prepared three representative compositions: an Hf-rich film with Hf/(Hf+Zr) ≈ 0.64, an intermediate film with Hf/(Hf+Zr) ≈ 0.50 (Hf:Zr ≈ 1:1), and a Zr-rich film with Hf/(Hf+Zr) ≈ 0.25. These three points sample the Hf-rich, intermediate, and Zr-rich regimes around the predicted design window. For each film, we measured the optical band gap and total dielectric constant. The optical band gap was obtained from transmission spectra using a Tauc analysis. The total dielectric constant was extracted from metal–insulator–metal capacitors after normalizing the capacitance by electrode area and film thickness.

Grazing-incidence X-ray diffraction in Figure 7a connects these compositions to the phase statistics in Figure 6b. For the Hf-rich film with Hf/(Hf+Zr) ≈ 0.64 (red curve), two strong peaks appear at about 28.5° and 31.5°, corresponding to the monoclinic (−111) and (111) reflections[35,36]. No pronounced peak is seen near 30.5°, indicating only minor tetragonal or orthorhombic content and confirming the suppression of high-symmetry phases[37]. At the intermediate composition Hf/(Hf+Zr) ≈ 0.50 (Hf:Zr ≈ 1:1, orange curve), the monoclinic (−111) and (111) peaks at 28.5° and 31.5° are strongly suppressed, while a sharp peak grows at about 30.5°. This peak matches the 111 reflections of the tetragonal and orthorhombic phases in $Hf_{1-x}Zr_xO_2$ and is intense and narrow. In other words, near Hf/(Hf+Zr) ≈ 0.50 the ALM films fall into the tetragonal–orthorhombic phase field that the thermodynamic map in Figure 6c identifies as a low-energy region. For the Zr-rich film with Hf/(Hf+Zr) ≈ 0.25 (green curve), the peak at 30.5° remains strong. The pattern is consistent with a tetragonal-dominated state, with no clear monoclinic signatures. The sequence from monoclinic-dominated diffraction at ≈ 0.64, through a mixed or orthorhombic-rich state at ≈ 0.50, to a tetragonal-dominated pattern at ≈ 0.25 follows the phase trend expected from the low-energy candidates in Figure 6b.

The electrical data in Figure 7b links this phase evolution to ferroelectric behavior. Ferroelectric switching behavior was assessed by referencing the P–E characteristics reported in our previous ALM study under an identical [Zr–Hf] sequence and comparable stack conditions. The Hf-rich sample, $Hf_{0.64}Zr_{0.36}O_2$, shows a narrow loop with small remanent polarization, consistent with a nonpolar monoclinic-dominated structure as in Figure 7a[38]. At the intermediate composition $Hf_{0.5}Zr_{0.5}O_2$ (Hf:Zr = 1:1), the loop becomes wide and well developed, and the remanent polarization reaches about $2P_r$ ≈ 35.1 μC cm$^{-2}$, typical for a ferroelectric film in which the polar orthorhombic phase is stabilized[39], validating the intended ferroelectric design. This is precisely the composition regime where the thermodynamic map predicts coexisting low-energy orthorhombic and tetragonal candidates. When the Zr content is increased to $Hf_{0.25}Zr_{0.75}O_2$, the loop evolves into a pinched or double-loop shape, characteristic of antiferroelectric-like behavior, and consistent with a tetragonal-dominated structure at Hf/(Hf+Zr) ≈ 0.25 [40].

The optical data in Figure 7c and Figure S3a are consistent with the predicted band gap trend and provide an independent check on the composition dependence of the electronic structure. The measured optical band gap decreases almost linearly from about 4.78 eV for the Hf-rich film to about 4.72 eV for the Zr-rich film as the Zr fraction increases. The change is modest, but both the sign and the magnitude of the slope agree well with the ALIGNN prediction, which also indicates a gradual reduction of the band gap toward the Zr-rich side. Because the film thickness is only about 4.5 nm, the extracted optical band gaps are slightly smaller than those reported for thicker reference films, which is reasonable given the stronger influence of interface states and sub-band edge absorption in such ultrathin layers[41].

Figure 7d and Figure S3b compares the measured total dielectric constant with the predicted electronic dielectric constant. The experimental values (red stars) exhibit a distinct dependence on the cation ratio, increasing significantly from 18.5 for the Hf-rich film to 29.1 at the equimolar composition, and peaking at 31.2 for the Zr-rich film (Hf/(Hf+Zr) = 0.25). Notably, this compositional trend is qualitatively captured by the theoretical prediction. The calculated electronic dielectric constant (blue squares) also displays an overall upward trajectory with a local maximum appearing near the Hf/(Hf+Zr) = 0.25 region, coinciding with the highest experimental value. Although the exclusion of ionic polarizability results in a quantitative offset, the IDEAL model successfully reproduces the experimental dielectric maximum in the Zr-rich regime. This convergence in compositional trends confirms that our framework accurately captures the intrinsic polarizability evolution governing the Hf–Zr–O system.

In summary, the ALM experiments and the P–E curves show a simple but consistent progression. Hf-rich films at Hf/(Hf+Zr) ≈ 0.64 are monoclinic-dominated, weakly polar, have the largest band gap, and the lowest dielectric constant; films near Hf/(Hf+Zr) ≈ 0.50 crystallize into orthorhombic- or tetragonal-like structures, show strong ferroelectric polarization, and keep a band gap only slightly reduced from the Hf-rich value. Zr-rich films are tetragonal-dominated, display antiferroelectric-like loops, possess the highest dielectric constant, and have the smallest band gap. This sequence agrees with phase statistics in Figure 6b and stability trend in Figure 6c, and it supports the model-suggested composition window as a practical region for combining phase stability, ferroelectric response, and band gap in ALM-grown $Hf_{1-x}Zr_xO_2$ films.

## Conclusions

We developed and demonstrated an inverse design framework that links large-scale foundation models for inorganic materials to ALD process design of $Hf_{1-x}Zr_xO_2$ thin films.

Starting from a fixed Hf, Zr, O chemical space, MatterGen generated a large and diverse set of candidate structures under a stability prior. CHGNet-based relaxation, together with a self-consistent $E_{hull}$ construction, provided a unified thermodynamic reference for both generated candidates and known Hf, Zr, O phases. By combining an $E_{hull}$ threshold of 0.1 eV atom$^{-1}$ with a stoichiometric window around a 1:2 cation to oxygen ratio, we reduced 10,000 generative samples to 991 thermodynamically and chemically plausible $Hf_{1-x}Zr_xO_2$-like structures.

ALIGNN predictions of band gap and electronic dielectric constant across this set, benchmarked against DFT and experimental references, reproduce the essential composition-dependent trends, while underestimating absolute values due to the DFT-based training data.

We translated these guidelines into experiments using ALM to deposit $Hf_{1-x}Zr_xO_2$ films with Hf/(Hf+Zr) ratios of approximately 0.64, 0.50, and 0.25. Grazing incidence X-ray diffraction confirmed the predicted progression from a monoclinic-dominated structure at Hf-rich composition, through a tetragonal and orthorhombic regime near $Hf_{0.5}Zr_{0.5}O_2$, to a tetragonal-dominated state at Zr-rich composition. Polarization–electric field (P–E) loops reported in our previous ALM study at comparable compositions show weak polarization for the Hf-rich film, strong ferroelectric response near $Hf_{0.5}Zr_{0.5}O_2$, and antiferroelectric-like behavior at Zr-rich composition, consistent with the phase statistics of the low energy candidates. In addition, the measured optical band gaps and dielectric constants follow the same composition-dependent trends as the ALIGNN predictions. The total dielectric constant increases from 18.5 in the Hf-rich film to 31.2 in the Zr-rich film, consistent with the increasing trend in the predicted electronic dielectric constant toward the Zr-rich region, although direct comparison of magnitudes is limited by the ionic contribution.

Overall, this IDEAL model shows that a generative model, a universal machine learning force field, and a graph neural network property predictor can be integrated into a practical inverse design platform that narrows the search space, identifies realistic composition and phase windows, and accelerates experimental optimization in complex oxide ALD. Because new generative models, machine learning interatomic potentials, and graph neural network property predictors continue to advance rapidly, the modular design of this platform allows straightforward replacement of the pretrained components, which can further improve accuracy and broaden the range of accessible process targets. We are currently extending this framework to the synthesis of new material systems beyond Hf, Zr, and O, and we will report these results in future work.

## Methods

**Building the IDEAL platform**

This study builds a model-driven platform that connects generative sampling of crystal structures, machine learning prediction of energies and properties, and comparison with known data for the Hf–Zr–O system. Because first-principles density functional theory (DFT) calculations are resource-intensive and difficult to scale to large generative candidate sets, we adopt a machine learning-based approach to enable high-throughput screening on a practical computational budget. The workflow has five main stages, as shown in Figure 1. First, MatterGen is used to generate a large set of hypothetical Hf–Zr–O crystal structures under a stability prior, as shown in Figure 1a.[8] Second, all generated structures are relaxed with the CHGNet in order to obtain consistent total energies and equilibrium geometries, as shown in Figure 1b.[10] Third, thermodynamic and compositional screening is applied using the $E_{hull}$ and a stoichiometric window around $Hf_{1-x}Zr_xO_2$, as shown in Figure 1c. Fourth, for the screened structures, a pre-trained ALIGNN model is used to estimate the band gap and electronic dielectric constant, and the resulting trends are compared with reported data[9], as shown in Figure 1d. Ultimately, as a final validation step, we translate these theoretical predictions into actual experimental protocols, as shown in Figure 1e. We perform ALM based on the model-suggested parameters to confirm the synthesizability and properties of the designed ternary oxides.

**Generative sampling with MatterGen**

MatterGen is a generative model for inorganic crystals that represents a structure by its lattice vectors and fractional atomic coordinates and learns the distribution of stable arrangements from large density functional theory datasets.[8] In generation mode, it starts from a random configuration, gradually denoises the lattice and atomic positions until a plausible structure that satisfies user-specified constraints is obtained, as shown in Figure 1a.

In this work, we use the 'chemical_system_energy_above_hull' generation mode of MatterGen, which is conditioned on both this chemical system and an approximate stability target, by requesting structures where the chemical system is fixed to Hf, Zr, and O, and the $E_{hull}$ is near 0.1 eV atom$^{-1}$ [4, 8]. Since the energy difference between the ground-state monoclinic phase and the ferroelectric orthorhombic phase in HfO$_2$-based systems is typically 0.05 eV atom$^{-1}$ [32, 42], the 0.1 eV atom$^{-1}$ threshold serves as a robust upper bound. It ensures the inclusion of these functional polymorphs while accounting for potential errors.

**Structure relaxation and energy evaluation with CHGNet**

The structures generated by MatterGen are not at an energy minimum. To obtain consistent energies and equilibrium geometries, we relax all candidates with CHGNet, as shown in Figure 1b. CHGNet is a graph neural network-based machine learning interatomic potential that models the potential energy surface of inorganic materials and is pre-trained on the Materials Project Trajectory Dataset, which contains more than ten years of DFT calculations of over 1.5 million inorganic structures, including energies, forces, stresses, and magnetic moments[10]. This large DFT dataset enables CHGNet to act as a universal potential for a wide range of chemistries at a cost much lower than DFT calculations.

Structures generated by MatterGen are used as the initial configuration for relaxation under CHGNet. Both atomic positions and lattice parameters are allowed to change. The relaxation proceeds in a series of geometry updates until the maximum atomic force falls below a chosen threshold and the residual stress is small. In this work, the force tolerance is set to about 0.02 eV Å$^{-1}$ [10]. The result of this stage is a set of relaxed Hf–Zr–O structures with CHGNet total energies that are suitable for thermodynamic analysis and property prediction.

**Thermodynamic and stoichiometric screening**

To verify the thermodynamic stability of the generated structures, we computed the $E_{hull}$ using a self-consistent CHGNet framework. We retrieved the structures of all stable and metastable reference phases in the Hf–O, Zr–O, and Hf–Zr–O systems from the Materials Project database via the pymatgen interface[43, 44], as shown in Figure 1b. These reference structures were then re-relaxed using CHGNet to obtain their formation energies on the same potential energy surface as the generated structures. $E_{hull}$ was computed as the energy difference between a candidate's formation energy and the energy of the convex hull at the same composition, using a self-consistent CHGNet relaxed reference set. This procedure eliminates systematic offsets between DFT data and machine learning predictions. We applied a stability threshold, $E_{hull}$, of 0.1 eV atom$^{-1}$, consistent with the generation target described in Section 2.2 and Figure 1c. Candidates exceeding this limit were discarded as thermodynamically unstable. Finally, we retained only structures with a cation-to-oxygen ratio (M:O) of approximately 1:2. This produces a final dataset that preserves the stoichiometry of the $Hf_{1-x}Zr_xO_2$ family while covering a diverse range of Hf-to-Zr ratios and structural motifs for subsequent property analysis.

**Property prediction with ALIGNN**

To estimate functional properties for the $Hf_{1-x}Zr_xO_2$-like candidates, we employ a pre-trained property model based on the ALIGNN architecture, as shown in Figure 1d. ALIGNN is a graph neural network that augments the standard atomistic graph with a line graph describing bonds, which allows angular interactions to be represented more accurately, and has been trained on large DFT datasets such as JARVIS DFT to predict electronic band gaps and electronic dielectric constant for inorganic crystals[9].

Each relaxed $Hf_{1-x}Zr_xO_2$ structure is passed to the ALIGNN model. We used the pretrained ALIGNN model for the electronic (clamped ion) dielectric constant and band gap as the primary screening target. We obtain predicted band gaps and electronic dielectric constants for all candidates in the filtered set. These predictions are used to construct distributions of properties as a function of composition and structural motif and to identify regions of composition space where a high dielectric constant and an acceptable band gap are simultaneously expected. In this work, the band gap and electronic dielectric constant are therefore used primarily to analyze trends and rank candidates, rather than to provide definitive values for individual structures.

**Experimental setup**

To experimentally validate the composition window identified by the computational platform, $Hf_{1-x}Zr_xO_2$ thin films were fabricated using ALM, following our previous work.[34, 45] Unlike conventional ALD, which typically realizes ternary compositions through super-cycles of binary $HfO_2$ and $ZrO_2$[29, 30], ALM exploits differences in steric hindrance between Hf and Zr precursors within a single deposition cycle to achieve atomic-level control of the cation ratio[34], as shown in Figure S1. To implement this, we employed a specific sequence designated as the [Zr–Hf] sequence (Zr precursor exposure, purge, Hf precursor exposure, purge, $O_3$ exposure, purge). As demonstrated in our previous study, this approach effectively yields superior film homogeneity and ferroelectric properties in the sub-10 nm thickness regime.[34]

The $Hf_{1-x}Zr_xO_2$ thin films were deposited in a commercial traveling-wave-type ALD reactor at a substrate temperature of 275 °C. Tetrakis(dimethylamido)hafnium (TDMAH) and tetrakis(ethylmethylamido)zirconium (TEMAZ) were used as the Hf and Zr precursors, respectively, and ozone ($O_3$, 120 g m$^{-3}$) served as the oxidant. Leveraging this established process, we finely tuned the precursor exposure times within the [Zr–Hf] sequence to synthesize thin films with three representative compositions: Hf-rich (Hf/(Hf+Zr) ≈ 64%), near-stoichiometric (Hf/(Hf+Zr) ≈ 50%), and Zr-rich (Hf/(Hf+Zr) ≈ 25%). These specific compositions were selected to validate the model predictions across distinct regions of the Hf–Zr–O phase diagram. For ferroelectric characterization, W/HfZrO/TiN capacitors were crystallized by post metallization annealing at 600 °C for 30 s in $N_2$, and P–E curves were measured using bipolar triangular pulses (pulse height of ±2 V at 1 kHz for 4.5 nm films).

## Data Availability

The data that support the findings of this study are available from the corresponding author upon reasonable request.

## Code Availability

The computer code used to generate the results in this study is available at

IDEAL (https://github.com/Gubonwook/IDEAL-Platform).

The IDEAL platform integrates the following open-source packages:

MatterGen (https://github.com/microsoft/mattergen).

CHGNet (https://github.com/CederGroupHub/chgnet).

ALIGNN (https://github.com/usnistgov/alignn).

## Acknowledgements

This work was conducted by the research grants (IRIS, NRF-2023M3H4A6A01057927) and (K-CHIPS, 2410012107, RS-2025-02311098, 25074-15FC) from the National Research

Foundation (NRF) funded by the Ministry of Science and ICT and the Ministry of Trade, Industry & Energy (MOTIE).## References

1. Parsons, G. N. & Clark, R. D. Area-Selective Deposition: Fundamentals, Applications, and Future Outlook. *Chem. Mater.* **32**, 4920–4953 (2020).

2. Mackus, A. J. M., Merkx, M. J. M. & Kessels, W. M. M. From the Bottom-Up: Toward Area-Selective Atomic Layer Deposition with High Selectivity †. *Chem. Mater.* **31**, 2–12 (2019).

3. Merchant, A. *et al.* Scaling deep learning for materials discovery. *Nature* **624**, 80–85 (2023).

4. Parida, C., Roy, D., Lastra, J. M. G. & Bhowmik, A. Mining Chemical Space with Generative Models for Battery Materials. *Batter. Supercaps* **202500309**, 1–10 (2025).

5. Han, N. & Su, B. L. AI-driven material discovery for energy, catalysis and sustainability. *Natl. Sci. Rev.* **12**, (2025).

6. Umeda, Y., Hayashi, H., Moriwake, H. & Tanaka, I. Prediction of dielectric constants using a combination of first principles calculations and machine learning. *Jpn. J. Appl. Phys.* **58**, 1–5 (2019).

7. Petousis, I. *et al.* Data Descriptor: High-throughput screening of inorganic compounds for the discovery of novel dielectric and optical materials. *Sci. Data* **4**, 1–12 (2017).

8. Zeni, C. *et al.* A generative model for inorganic materials design. *Nature* **639**, 624–632 (2025).

9. Choudhary, K. & DeCost, B. Atomistic Line Graph Neural Network for improved materials property predictions. *npj Comput. Mater.* **7**, 1–8 (2021).
29

10. Deng, B. *et al.* CHGNet as a pretrained universal neural network potential for charge-informed atomistic modelling. *Nat. Mach. Intell.* **5**, 1031–1041 (2023).

11. Sneh, O., Clark-Phelps, R. B., Londergan, A. R., Winkler, J. & Seidel, T. E. Thin film atomic layer deposition equipment for semiconductor processing. *Thin Solid Films* **402**, 248–261 (2002).

12. Hautier, G., Ong, S. P., Jain, A., Moore, C. J. & Ceder, G. Accuracy of density functional theory in predicting formation energies of ternary oxides from binary oxides and its implication on phase stability. *Phys. Rev. B - Condens. Matter Mater. Phys.* **85**, (2012).

13. Bartel, C. J. *et al.* A critical examination of compound stability predictions from machine-learned formation energies. *npj Comput. Mater.* **6**, 1–11 (2020).

14. Greeley, J., Jaramillo, T. F., Bonde, J., Chorkendorff, I. & Nørskov, J. K. Computational high-throughput screening of electrocatalytic materials for hydrogen evolution. *Nat. Mater.* **5**, 909–913 (2006).

15. Curtarolo, S. *et al.* The high-throughput highway to computational materials design. *Nat. Mater.* **12**, 191–201 (2013).

16. Takahashi, A., Kumagai, Y., Miyamoto, J., Mochizuki, Y. & Oba, F. Machine learning models for predicting the dielectric constants of oxides based on high-throughput first-principles calculations. *Phys. Rev. Mater.* **4**, 103801 (2020).

17. Karsch, F., Patkós, A. & Petreczky, P. Phonons and related crystal properties from density-functional perturbation theory. *Phys. Lett. Sect. B Nucl. Elem. Part. High-Energy Phys.* **401**, 69–73 (1997).

18. Gonze, X. & Lee, C. Dynamical matrices, Born effective charges, dielectric permittivity tensors, and interatomic force constants from density-functional perturbation theory. *Phys. Rev. B - Condens. Matter Mater. Phys.* **55**, 10355–10368 (1997).

19. Choudhary, K. *et al.* The joint automated repository for various integrated simulations (JARVIS) for data-driven materials design. *npj Comput. Mater.* **6**, (2020).

20. Chadi, D. J. & White, R. M. Frequency- and wave-number-dependent dielectric function of semiconductors. *Phys. Rev. B* **11**, 5077–5081 (1975).

21. Shannon, R. D. Dielectric polarizabilities of ions in oxides and fluorides. *J. Appl. Phys.* **73**, 348–366 (1993).

22. Luo, X., Zhou, W., Ushakov, S. V., Navrotsky, A. & Demkov, A. A. Monoclinic to tetragonal transformations in hafnia and zirconia: A combined calorimetric and density functional study. *Phys. Rev. B - Condens. Matter Mater. Phys.* **80**, 1–13 (2009).

23. Rignanese, G. M. Dielectric properties of crystalline and amorphous transition metal oxides and silicates as potential high-κ candidates: The contribution of density-functional theory. *J. Phys. Condens. Matter* **17**, 357–379 (2005).

24. Zhao, X. & Vanderbilt, D. First-principles study of structural, vibrational, and lattice dielectric properties of hafnium oxide. *Phys. Rev. B - Condens. Matter Mater. Phys.* **65**, 1–4 (2002).

25. Zheng, J. X., Ceder, G., Maxisch, T., Chim, W. K. & Choi, W. K. First-principles study of native point defects in hafnia and zirconia. *Phys. Rev. B - Condens. Matter Mater. Phys.* **75**, 1–7 (2007).

26. Cohen, A. J., Mori-Sánchez, P. & Yang, W. Challenges for density functional theory. *Chem. Rev.* **112**, 289–320 (2012).

27. Robertson, J. Band offsets of wide-band-gap oxides and implications for future electronic devices. *J. Vac. Sci. Technol. B Microelectron. Nanom. Struct. Process. Meas. Phenom.* **18**, 1785–1791 (2000).

28. Horton, M. K., Dwaraknath, S. & Persson, K. A. Promises and perils of computational materials databases. *Nat. Comput. Sci.* **1**, 3–5 (2021).

29. Müller, J. *et al.* Ferroelectricity in simple binary ZrO 2 and HfO 2. *Nano Lett.* **12**, 4318–4323 (2012).

30. Hyuk Park, M. *et al.* Evolution of phases and ferroelectric properties of thin Hf 0.5Zr0.5O2 films according to the thickness and annealing temperature. *Appl. Phys. Lett.* **102**, 0–5 (2013).

31. Park, M. H. *et al.* Ferroelectricity and Antiferroelectricity of Doped Thin HfO2-Based Films. *Adv. Mater.* **27**, 1811–1831 (2015).

32. Materlik, R., Kunneth, C. & Kersch, A. The origin of ferroelectricity in Hf1-xZrxO2: A computational investigation and a surface energy model. *J. Appl. Phys.* **117**, (2015).

33. Song, T. *et al.* Stabilization of the Ferroelectric Phase in Epitaxial Hf1-xZrxO2 Enabling Coexistence of Ferroelectric and Enhanced Piezoelectric Properties. *ACS Appl. Electron. Mater.* **3**, 2106–2113 (2021).

34. Trinh, N. Le *et al.* Atomic-Level Stoichiometry Control of Ferroelectric HfxZryOz Thin Films by Understanding Molecular-Level Chemical Physical Reactions. *ACS Nano* **19**, 3562–3578 (2025).

35. Li, Y. *et al.* A Ferroelectric Thin Film Transistor Based on Annealing-Free HfZrO Film. *IEEE J. Electron Devices Soc.* **5**, 378–383 (2017).

36. Yang, X. *et al.* Composition-dependent structure and bandgaps in HfxZr1−xO2 thin films. *Appl. Phys. Lett.* **124**, 1–6 (2024).

37. Triyoso, D. H., Gregory, R., Park, M., Wang, K. & Lee, S. I. Physical and Electrical Properties of Atomic-Layer-Deposited Hf[sub x]Zr[sub 1−x]O[sub 2] with TEMAHf, TEMAZr, and Ozone. *J. Electrochem. Soc.* **155**, H43 (2008).

38. Dou, X. *et al.* Polarization switching pathways of ferroelectric Zr-doped HfO2 based on the first-principles calculation. *Appl. Phys. Lett.* **124**, 0–5 (2024).

39. Kim, S. J. *et al.* Effect of film thickness on the ferroelectric and dielectric properties of low-temperature (400 °C) Hf0.5Zr0.5O2 films. *Appl. Phys. Lett.* **112**, 095045 (2018).

40. Goh, Y., Hwang, J. & Jeon, S. Excellent Reliability and High-Speed Antiferroelectric HfZrO2Tunnel Junction by a High-Pressure Annealing Process and Built-In Bias Engineering. *ACS Appl. Mater. Interfaces* **12**, 57539–57546 (2020).

41. Zhang, K. H. L. *et al.* Thickness dependence of the strain, band gap and transport properties of epitaxial In2O3 thin films grown on Y-stabilised ZrO 2(111). *J. Phys. Condens. Matter* **23**, (2011).

42. Huan, T. D., Sharma, V., Rossetti, G. A. & Ramprasad, R. Pathways towards ferroelectricity in hafnia. *Phys. Rev. B - Condens. Matter Mater. Phys.* **90**, 1–5 (2014).

43. Jain, A. *et al.* Commentary: The materials project: A materials genome approach to accelerating materials innovation. *APL Mater.* **1**, (2013).

44. Ong, S. P. *et al.* Python Materials Genomics (pymatgen): A robust, open-source python library for materials analysis. *Comput. Mater. Sci.* **68**, 314–319 (2013).

45. Nguyen, C. T. *et al.* Atomic Layer Modulation of Multicomponent Thin Films through Combination of Experimental and Theoretical Approaches. *Chem. Mater.* **33**, 4435–4444 (2021).

## Graphical Abstract

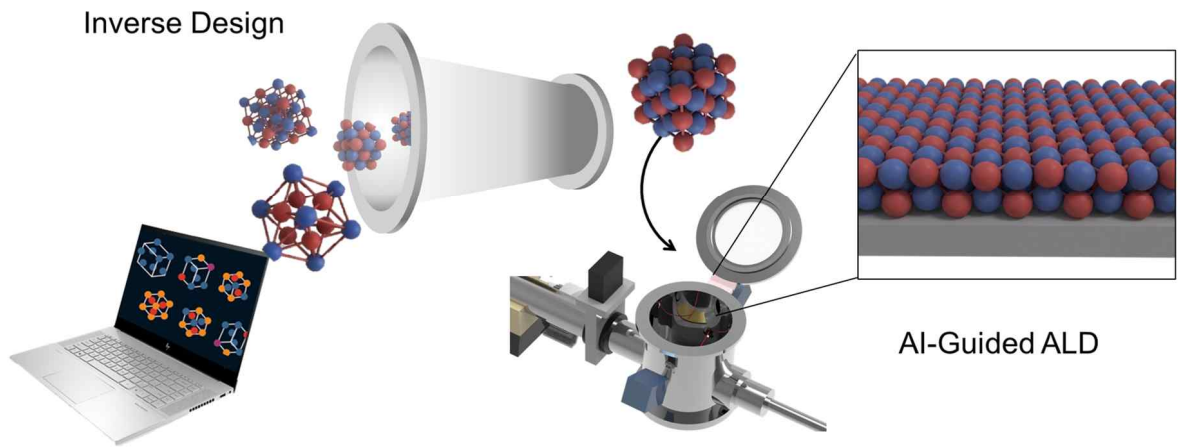

The IDEAL platform connects generative foundation models with atomic layer deposition to minimize trial-and-error in materials discovery. Using the Hf–Zr–O system as a benchmark, the framework successfully predicts a thermodynamic window for stable ferroelectric phases. These predictions are experimentally validated, establishing a generalizable pathway for the inverse design of functional semiconductor thin films.